%% file: main.tex
\documentclass[sigconf]{acmart}

\AtBeginDocument{%
  \providecommand\BibTeX{{%
    \normalfont B\kern-0.5em{\scshape i\kern-0.25em b}\kern-0.8em\TeX}}}

\setcopyright{acmlicensed}
\copyrightyear{2024}
\acmYear{2024}

\acmConference[EDM LLM Workshop '24]{EDM LLM Workshop}{July
  2024}{Atlanta, GA}
\usepackage{multirow}

\begin{document}

\title[Safe Generative Chats]{Safe Generative Chats in a WhatsApp Intelligent Tutoring System}

\author{Zachary Levonian}
\email{zach@levi.digitalharbor.org}
\orcid{0000-0002-8932-1489}
\affiliation{%
  \institution{Digital Harbor Foundation}
  \streetaddress{1045 Light St.}
  \city{Baltimore}
  \state{Maryland}
  \country{USA}
  \postcode{21230}
}

\author{Owen Henkel}
\email{owen.henkel@education.ox.ac.uk}
\affiliation{%
  \institution{University of Oxford}
  \country{UK}
}

\renewcommand{\shortauthors}{Levonian et al.}
\newcommand{\staremoji}{\includegraphics[height=7pt]{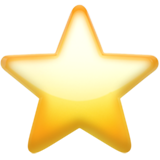}}

\begin{abstract}
Large language models (LLMs) are flexible, personalizable, and available, which makes their use within Intelligent Tutoring Systems (ITSs) appealing.
However, that flexibility creates risks: inaccuracies, harmful content, and non-curricular material.
Ethically deploying LLM-backed ITS systems requires designing safeguards that ensure positive experiences for students.
We describe the design of a conversational system integrated into an ITS, and our experience evaluating its safety with red-teaming, an in-classroom usability test, and field deployment.
We present empirical data from more than 8,000 student conversations with this system, finding that GPT-3.5 rarely generates inappropriate messages.
Comparatively more common is inappropriate messages from students, which prompts us to reason about safeguarding as a content moderation and classroom management problem.
The student interaction behaviors we observe provide implications for designers---to focus on student inputs as a content moderation problem---and implications for researchers---to focus on subtle forms of bad content.
\end{abstract}



\received{10 May 2024}
\received[accepted]{31 May 2024}

\maketitle

\input{body}

\begin{acks}
We would like to thank Millie-Ellen Postle, Hannah Horne-Robinson, and the staff of Rising Academies for their contributions.
This work was supported by the Learning Engineering Virtual Institute (LEVI) and
the Digital Harbor Foundation.
\end{acks}

\bibliographystyle{ACM-Reference-Format}
\bibliography{levi_enghub_zotero}



\end{document}

%% file: body.tex
\section{Introduction}




The capabilities of Large Language Models (LLMs) have led to a surge of interest in applying them to educational settings, including for automated tutoring, personalized learning, and adaptive assessment~\cite{kasneci_chatgpt_2023,caines_application_2023}. 
A particularly promising application of LLMs is integration with Intelligent Tutoring Systems (ITSs), as they can combine the structured pedagogical processes and vetted curricula of ITSs and the flexibility and personalization enabled by conversational interfaces~\cite{levonian_retrieval-augmented_2023,upadhyay_improving_2023,hobert_say_2019,khosrawi-rad_conversational_2022}.

Integrating LLMs into ITSs enables answering student questions, summarizing concepts, creating customized hints, and recontextualizing learning materials~\cite{pardos_learning_2023,sonkar_class_2023,levonian_retrieval-augmented_2023}.
However, the use of LLMs in educational applications also raises concerns regarding potential risks, including the generation of toxic language, implicit biases, and inaccurate information, as well as inappropriate use by students~\cite{liang_towards_2021,navigli_biases_2023,baker_algorithmic_2022,yan_practical_2024}.
These risks become particularly important when designing educational applications that directly interact with students e.g. via a chat interface, necessitating a focus on the safety and accuracy of model-generated responses to students.

Recent advancements in LLMs have led to improvements in mitigating some of the most distressing behaviors of early generations, such as toxicity, wildly inaccurate information, and discussions of illegal or taboo topics~\cite{openai_gpt-4_2023,tao_auditing_2023}.
While this progress is welcome, it has also revealed a range of more subtle potential problems. 
For instance, small hallucinations (e.g., confusing ``$2\pi $r'' with ``$\pi r^2$'') may lead to persistent misconceptions~\cite{murphy_situating_2013}.
Additionally, younger students might be more likely to anthropomorphize models and develop emotionally charged relationships with them~\cite{girard_what_2010,goldman_preschoolers_2023}, and models tend to present an ``average'' view of the Anglophone internet which might not be appropriate in certain cultural contexts~\cite{xu_survey_2024,agiza_analyzing_2024}.

Less discussed is how models should handle inappropriate or potentially offensive student inputs, as well as honest questions on politically or culturally sensitive topics. For example, if a student addresses an LLM application using profane language, should the model ignore the profanity and proceed, ask the student to stop using such language, or request that the student rephrase the question? Similarly, if a student asks an honest question about a potentially charged political topic (e.g., ``Is it okay to get pregnant before you are married?''), should the model provide a standard ``it depends'' answer, ignore the question, or inform the student that they cannot discuss the topic?

Perhaps most seriously, if a student discloses some sort of trauma or abuse they have suffered, how should the model respond? While these are complex questions, they are also ones that teachers and tutors deal with regularly~\cite{falkiner_teachers_2017,beck_knowledge_1994,haney_ethical_2004,berger_school_2022}. 
Deciding how to respond to inappropriate or provocative student questions is a classic challenge of classroom management~\cite{sabornie_handbook_2022,marzano_handbook_2005}, carefully choosing how to address and explain sensitive topics is a fraught area for nearly all teachers~\cite{levin_what_2002,falkiner_heads_2020}, and handling sensitive student disclosures is such an important question that most school systems have codified mandatory reporting rules for teachers that specify which types of student disclosures must be reported to school leadership, mental health professionals, or law enforcement~\cite{goldman_primary_2007}.

In this paper, we describe a system we designed for safeguarding student chats with an ITS and empirical data from a field deployment of that system with usage from more than 8,000 students. 
We formed a research collaboration with the developers of Rori, a WhatsApp-based chatbot math tutor.
Rori is used primarily by low-income middle-school students in Sierra Leone, Liberia, Ghana, and Rwanda both in classroom settings and at home for math skills practice~\cite{henkel_effective_2024}.
We designed a conversational experience for Rori's users that teaches them about growth mindset before they begin math skill practice.

Dinan et al. identify three broad safety issues in conversational systems: (a) instigator effects, in which the system generates harmful content, (b) yea-sayer effects, in which the system endorses or fails to object to harmful content, and (c) imposter effects, in which the system provides incorrect or harmful advice~\cite{dinan_anticipating_2021}. 
To safeguard students during that conversation, we designed a safety system consisting of filters and corresponding actions when messages are flagged by those filters. Across two studies, we find no evidence of instigator or imposter effects but limited evidence of the yea-sayer effect.

In study one, we assess the usability and ethical acceptability of the system in the classroom.
In study two, we deploy the system for use by students at home.
The empirical evidence from these two studies provides implications for designers---to focus on student inputs as a content moderation problem---and implications for researchers---to focus on subtle forms of bad content.

\section{System Design}

\begin{figure}[t]
  \centering
  \includegraphics[width=\linewidth]{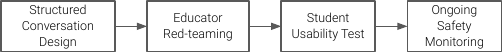}
  \caption{Designing for safety: our process.}
  \Description{A flowchart of our safety design process.}
  \label{fig:design_flowchart}
\end{figure}

\begin{figure}[h]
  \centering
  \includegraphics[width=\linewidth]{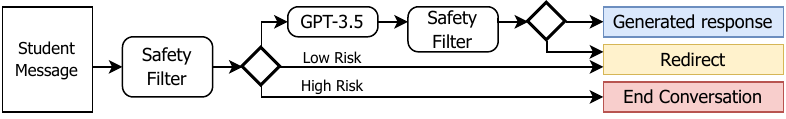}
  \caption{The generative chat moderation system.}
  \Description{A flowchart of our moderation process.}
  \label{fig:moderation_flowchart}
\end{figure}

\begin{figure}[t]
  \centering
  \includegraphics[width=\linewidth]{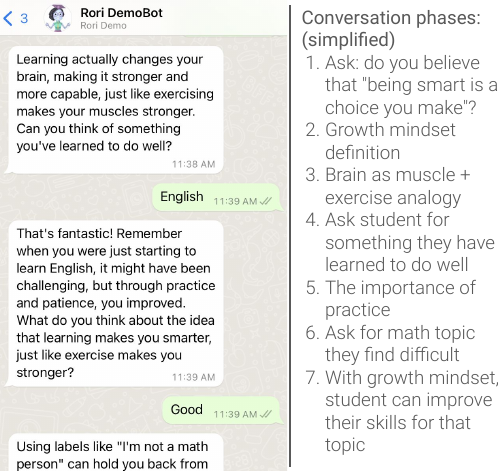}
  \caption{A chat excerpt from the Rori WhatsApp interface and a simplified view of the conversation phases.}
  \Description{A simplified overview of the seven conversational phases in the growth mindset conversation.}
  \label{fig:interface}
\end{figure}

To design a safe generative chat experience, we implemented a system on the basis of educator feedback and through multiple phases of evaluation as shown in Figure~\ref{fig:design_flowchart}.

\textbf{Designing a semi-structured conversation.}
We chose to implement a generative chat for encouraging a growth mindset, an approach linked to positive educational outcomes, including in mobile learning contexts~\cite{karumbaiah_addressing_2017,yeager_using_2016,kizilcec_growth_2019}.
We used a prompting approach that moves the conversation through multiple phases: introducing the concept of a growth mindset, asking the student to reflect on a time that practice has helped them, and identifying a specific math skill that they want to practice.
The system initiates the conversation with the message ``Do you agree with the statement `Being smart is a choice you make, not the way you are'?'' and moves the student through various conversational phases, as shown in Figure~\ref{fig:interface}. 
During the conversation, we detect standard navigation keywords (e.g. ``menu'') to navigate away from the conversation and on to math skills practice. 
We limited the total conversation length---a max of 8 turns during the usability test and 10 during the field deployment---to decrease the chance of major digressions and to reduce any student frustration.
By designing the conversation as system-initiated rather than student-initiated and ending each system message with a question, we provide structure that keeps the conversation flowing and focused on growth mindset.

\textbf{Designing safety guardrails.}
To ensure students have a safe experience during the conversation, we implemented guardrails that would redirect or end the conversation.
Each student and system message is passed through a safety filter that determines how the system will respond to the student.
Figure \ref{fig:moderation_flowchart} demonstrates the final design.
The safety filter consists of (1) a word list and (2) a statistical moderation model.
The word list---consisting only of unambiguous curse words---is applied first. While a word list is rigid and inflexible, we chose to include it because it is easier for educators and parents to reason about than a statistical model~\cite{jhaver_personalizing_2023}.
The statistical model we used was OpenAI's moderation API, which predicts the presence of five high-level content categories and six sub-categories~\cite{noauthor_moderation_2024}.
Each message is given a score between 0 and 1 reflecting how likely that message is to contain content in that category.
We set the per-category thresholds for which we would take system action based on the red-teaming exercise.

\textbf{System moderation actions.}
Based on the assessed risk of the message, we took one of two moderation actions in response to student messages.
We classified self-harm, sexual/minors, and the two /threatening sub-categories as high risk messages and the rest as low risk. 
In response to low risk messages, we drop the students most recent message from the prompted context and ask them to continue the conversation with a more appropriate message.
In response to high risk messages, we end the conversation immediately with the message: ``That sounds like a serious topic, and a real person needs to look at this.  They might try to contact you to check on you. Until someone has reviewed this, Rori will not reply.''
We make an open source reference implementation of our moderation system available on GitHub.\footnote{\url{https://github.com/DigitalHarborFoundation/chatbot-safety}}

\textbf{Educator red-teaming.} To evaluate the acceptability of the conversation design and the safety guardrails, we conducted an asynchronous red-teaming exercise.
There is considerable variation in red-teaming exercises~\cite{feffer_red-teaming_2024}; the purpose of our exercise was to qualitatively assess the effectiveness of the safety guardrails and to quantitatively set initial per-category moderation thresholds.
We recruited 17 Rising Academies educators and system designers to adversarially probe the conversation design.
Across 57 conversations, we received negative feedback on 39 messages that should have been flagged, setting the thresholds appropriately. After small tweaks to the prompts, we observed no obviously negative conversational experiences.
We return to the topic of subtly negative experiences in the discussion, but we determined there to be minimal risk in proceeding with a full usability assessment with students.

\textbf{Monitoring.}
To ensure the safety of Rori student users, we designed a continual monitoring procedure.
We implemented data dashboards to review the most recent and the riskiest conversations. 
Messages flagged as high-risk generate an email alert to an internal team.
We designed a basic reporting protocol for use with student users in the event of particular sensitive disclosures e.g. sexual abuse or suicidal thoughts.

\begin{table}[t]
\caption{Counts of students, conversations, and messages across two studies.}
\begin{tabular}{@{}lrrr@{}}
\toprule
                 & Students & Conversations & Messages \\ \midrule
Usability Test   & 109 & 252 & 3,722 \\
Field Deployment & 8,168 & 8,755 & 126,278 \\ \bottomrule
\end{tabular}
\label{tab:participant_counts}
\end{table}

\section{Study 1: Student Usability Test}

\begin{figure}[htb]
  \centering
  \includegraphics[width=\linewidth]{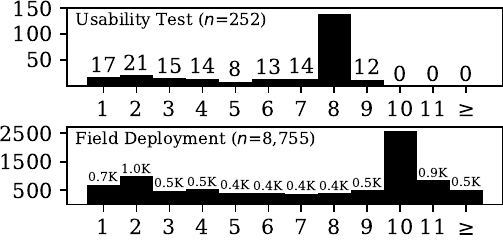}
  \caption{Conversation length (as number of student messages) for all conversations. Completion rate was higher during the usability test (59.5\%) than the field deployment (38.9\%).}
  \Description{Histograms showing conversation length; distributions are mostly uniform, with large spikes at the max conversation length (8 during the usability test, 10 during the field deployment).}
  \label{fig:conversation_length}
\end{figure}

\begin{table}[htb]
\caption{Highest and 99th percentile of the OpenAI moderation scores observed during the two studies. The highest possible value is 1.}
\begin{tabular}{@{}crrrr@{}}
\toprule
 & \multicolumn{2}{c}{Usability Test} & \multicolumn{2}{c}{Field Deployment} \\ 
Source & Q99 & Max & Q99 & Max \\ \midrule
GPT-3.5 & 0.000 & 0.010 & 0.003 & 0.044 \\ 
Student & 0.002 & 0.045 & 0.030 & 0.989 \\
\bottomrule
\end{tabular}
\label{tab:moderation_scores_overview}
\end{table}


\begin{table}[htb]
\caption{Student conversation ratings during Study 1.}
\begin{tabular}{@{}c|cccccc@{}}
\toprule
Rating             & none & \staremoji\staremoji\staremoji\staremoji\staremoji & \staremoji\staremoji\staremoji\staremoji & \staremoji\staremoji\staremoji & \staremoji\staremoji & \staremoji \\
\# conversations & 125       & 126         & 4           & 5           & 2           & 5           \\ \bottomrule
\end{tabular}
\label{tab:ratings}
\end{table}

In December 2023, 109 in-school students across 6 total classrooms were instructed to use the growth mindset generative chat during a regularly-scheduled study hall using Rori for math skills practice~\cite{henkel_effective_2024}.
252 conversations occurred between December 13th and 15th.
60\% of the conversations were completed; the distribution of conversation lengths is shown in Figure~\ref{fig:conversation_length}.\footnote{Due to a bug that under-counted student messages, some conversations continued an extra turn.}

At the end of the conversation, we asked students to rate the conversation from one to five stars.\footnote{Feedback request message: ``Thank you for your time! How much did you like the conversation?'' A response modal labeled ``Give us some \staremoji s!'' has quick-reply buttons.}
The distribution of ratings is shown in Table~\ref{tab:ratings}. Of the rated conversations, 16 conversations (6.3\%) were rated less than five stars.
Qualitative investigation of those 16 low-rated conversations reveals no clear difference between those and 5-star conversation; student messages in low-rated conversations were non-significantly more likely to be single-word responses (75.4\% low-rated vs 65.4\% five-star, $\chi^2$=0.82, d.f.=1, $p$=0.36).

No student or GPT-3.5 student messages were flagged by the safety filter.
In particular, most GPT-3.5 and student messages received low moderation scores across all categories. Table~\ref{tab:moderation_scores_overview} shows summary statistics for the highest score received across all categories: the highest-scoring GPT-3.5 message received a score of 0.01 (``Oh, it seems like you might not understand the question. Let me rephrase it. Do you think that being smart is something that you can choose to be, rather than something that you are born with?''), while the highest-scoring student message received a score of 0.05 (a typo).

It may be that the moderation API's implicit values diverge from our own, such that false negatives occur and harmful student messages are not flagged.
To check, we randomly sampled 100 student conversations, finding no false negatives. Qualitatively, while some student messages were playful or inappropriate in ways that would likely trigger a response from a human tutor, we found our prompt for GPT-3.5 effective at producing appropriate redirections back to the current topic.

Taken together, these results suggest that the semi-structured growth mindset conversation is acceptable for broader use.
Critically, the conversation design was effective at preventing messages that would trigger the safety filter: we identified no obviously unacceptable student messages.
We made minor adjustments to the prompts and proceeded to a field deployment.

\section{Study 2: Field Deployment}

\begin{table}[htb]
\caption{OpenAI moderation scores by category for the 54,384 student messages sent during the field deployment. In addition to the 99th percentile and maximum observed score over all student messages, we show the number of messages with a score greater than 0.1 and greater than 0.5.}
\begin{tabular}{@{}lrrrr@{}}
\toprule
Category & Q99 & Max & $n\ge0.1$ & $n\ge0.5$ \\ 
\midrule
Harassment & 0.011 & 0.989 & 141 & 36 \\
Sexual & 0.012 & 0.914 & 28 & 5 \\
Hate & 0.002 & 0.524 & 3 & 1 \\
Violence & 0.001 & 0.959 & 2 & 1 \\
Self-harm/intent & 0.001 & 0.743 & 1 & 1 \\
Self-harm & 0.001 & 0.531 & 1 & 1 \\
Harassment/threatening & 0.000 & 0.451 & 1 & 0 \\
Hate/threatening & 0.000 & 0.087 & 0 & 0 \\
Violence/graphic & 0.000 & 0.081 & 0 & 0 \\
Self-harm/instructions & 0.000 & 0.072 & 0 & 0 \\
Sexual/minors & 0.007 & 0.024 & 0 & 0 \\
\bottomrule
\end{tabular}
\label{tab:moderation_scores_by_category}
\end{table}

The growth mindset conversation was deployed publicly on Feburary 13, 2024 for non-school users of Rori and incorporated as a component of the on-boarding process before math skills practice begins.
We analyzed the 126,278 messages between the feature launch and May 1, 2024.

\subsection{Did GPT-3.5 generate objectionable outputs?}

No. Quantitatively, the highest-scoring system message produced received a score of 0.044.
During continual monitoring, the researchers annotated GPT-3.5 messages and determined none of them to be objectionable.
The most controversial messages were those generated in response to student's objectionable messages, which we discuss in the next sections.

\subsection{Did students write objectionable messages?}

Yes, but not very much. 
0.31\% of student messages received a score in any moderation category of at least 0.1.
Fewer than 8 in 10000 messages were flagged.
Table~\ref{tab:moderation_scores_by_category} summarizes the moderation scores per-category.
The most common negative messages were harassing or sexual.
Only one message was flagged as high risk.
After investigation by the team, it was determined to be a false positive by the OpenAI moderation model---the message should have been classified as low risk, as it contained violent language that merited corrective action but did not evidence self-harm.
From an investigation of the 27 conversations with flagged messages, all flagged messages were determined to merit corrective action.

\subsection{Did GPT-3.5 respond appropriately?}

We investigated the messages generated in response to student messages that were near the safety filter thresholds but remained unflagged.
48 unflagged conversations contained a message with a moderation score of at least 0.1. 
40 of these conversations included at least one student message that warranted caution or a corrective statement from the system response, and we deemed the GPT-3.5-generated responses to be appropriately corrective in 37 of those cases. 
In 3 cases, the generated response ignored or equivocated when a corrective message would have been warranted.
This is a subtler form of bad response: the yea-sayer effect~\cite{dinan_anticipating_2021}.

\section{Discussion}


\subsection{Key Findings}

In this workshop paper, we described a system for conducting safe generative chats inside of an existing ITS.
We found that the semi-structured conversation design we used eliminated imposter effects, while safety filters for students' inputs eliminated instigator effects.
We found that it was surprisingly straightforward to develop a prompt for GPT-3.5 to respond appropriately to the vast majority of student messages~\cite{narayanan_model_2023}.

Instead, our attention was drawn to the more frequent and more challenging problem of how to deal with inappropriate or otherwise sensitive student messages. In some ways this challenge in analogous to the challenges of content moderation on online platforms, where the context in which a comment exists is important, and policies that are reasonable in many cases might be ineffective in edge cases.  As an example: how to handle questions regarding contentious political or historical topics? In many cases acknowledging that there are different valid opinions is a good pedagogical approach, but in particularly sensitive or egregious examples this ``both-sideism'' can be inappropriate~\cite{smith_how_2022}. However, these are the types of challenges teachers deal with constantly, 
and we believe that there is a research opportunity here at the intersection of content moderation and classroom management to develop appropriate system actions in response to objectionable student messages.

Another important finding was that the process of red-teaming was effective in its primary goal of identifying potential risk. It had other benefits we did not expect: building organizational confidence.
We found that being transparent about the shortcoming of our V1 approach and including designers, educators, and researchers in the evaluation process had the dual benefit of improving trust and soliciting higher-quality feedback to improve the design. 

\subsection{Limitations \& Future Research}

The specific moderation actions we implemented are reasonable starting points, and by classifying messages at two risk levels we are able to positively redirect conversations with pre-vetted messages~\cite{leach_using_2016}.
While these corrective messages were written by educators, in the future we hope that approaches from culturally-responsive classroom management might be combined with soliciting cultural background information from students so that behavioral expectations can be communicated more clearly and correctives can be applied more appropriately~\cite{weinstein_toward_2004,skiba_teaching_2016}.

In the event of more serious disclosures, as with the messages we classify as high risk, we argue that our choice to automatically end the conversation and move to human review rather than attempting to generate an appropriate LLM response in the moment is the more ethical one~\cite{stapleton_if_2024}.
However, the specific approach we used of ending the conversation is not ideal; we might consider technical infrastructure that starts an in-chat support session with a human or otherwise connects explicitly to contacts at the student's school.

We did observe evidence of the yea-sayer effect in response to some objectionable student messages; future work should explore opportunities for mitigating this effect.
In the mean time, designers should monitor for the prevalence of yea-saying and consider technical approaches that explicitly model the appropriate corrective behavior.


